\documentclass{article}

\usepackage[final]{neurips_2021_ml4ps}

\usepackage[utf8]{inputenc} 
\usepackage[T1]{fontenc}    
\usepackage{hyperref}       
\usepackage{url}            
\usepackage{booktabs}       
\usepackage{amsfonts}       
\usepackage{nicefrac}       
\usepackage{microtype}      
\usepackage[dvipsnames]{xcolor}         

\usepackage[labelformat=parens,labelsep=quad, skip=3pt]{caption}
\usepackage{graphicx}
\usepackage{subfloat}
\usepackage[subrefformat=parens]{subcaption}
\usepackage{amsmath}
\usepackage[nolist]{acronym}

\let\svthefootnote\thefootnote
\newcommand\freefootnote[1]{%
  \let\thefootnote\relax%
  \footnotetext{#1}%
  \let\thefootnote\svthefootnote%
}

\begin{acronym}[ELAIS-S1]
\acro{2dfGRS}{2dF Galaxy Redshift Survey}
\acro{AGN}{Active Galactic Nucleus}
\acroplural{AGN}{Active Galactic Nucleii}
\acro{ASKAP}{Australian Square Kilometre Array Pathfinder}
\acro{ATCA}{Australia Telescope Compact Array}
\acro{ATLAS}{Australia Telescope Large Area Survey}
\acro{COSMOS}{COSMic evOlution Survey}
\acro{DES}{Dark Energy Survey}
\acro{DT}{Decision Tree}
\acro{eCDFS}{extended Chandra Deep Field South}
\acro{ELAIS-S1}{European Large Area ISO Survey--South 1}
\acro{EMU}{Evolutionary Map of the Universe}
\acro{GP}{Gaussian Process}
\acroplural{GP}{Gaussian Processes}
\acro{kNN}{$k$-Nearest Neighbours}
\acro{LMNN}{Large Margin Nearest Neighbour}
\acro{LOFAR}{LOw Frequency ARray}
\acro{LOTSS}{LOFAR Two-metre Sky Survey}
\acro{ML}{Machine Learning}
\acro{MLKR}{Metric Learning for Kernel Regression}
\acro{MOS}{Multi-Object Spectroscopy}
\acro{MSE}{Mean Square Error}
\acro{MWA}{Murchison Widefield Array}
\acro{NMAD}{Normalised Median Absolute Deviation}
\acro{NMI}{Normalised Mutual Information}
\acro{OzDES}{Australian Dark Energy Survey}
\acro{QSO}{Quasi-Stellar Object}
\acro{RF}{Random Forest}
\acro{SDSS}{Sloan Digital Sky Survey}
\acro{SED}{Spectral Energy Distribution}
\acro{SKA}{Square Kilometre Array}
\acro{SKADS}{Square Kilometre Array Design Survey}
\acro{SST}{Spitzer Space Telescope}
\acro{SWIRE}{Spitzer Wide-Area Infrared Extragalactic Survey}
\acro{WAVES}{Wide Area Vista Extragalactic Survey}
\acro{GAN}{Generative Adversarial Network}
\acro{GAIN}{Generative Adversarial Imputation Network}
\acro{MICE}{Multivariate Imputation by using Chained Equation}
\acro{RMSE}{Root Mean Square Error}
\acro{NMAD}{Normalised Median Absolute Deviation}
\end{acronym}

\title{Missing Data Imputation for Galaxy Redshift Estimation}

\author{%
  Kieran J. Luken$^*$, \quad Rabina Padhy, \quad X. Rosalind Wang$^*$ \\
  Western Sydney University, 
  Penrith, NSW, Australia\\
  \texttt{k.luken@westernsydney.edu.au} \\
  \texttt{rosalind.wang@westernsydney.edu.au} \\
}

\begin{document}

\maketitle

\begin{abstract}
  Astronomical data is full of holes. While there are many reasons for this missing data, the data can be randomly missing, caused by things like data corruptions or unfavourable observing conditions. We test some simple data imputation methods (Mean, Median, Minimum, Maximum and \ac{kNN}), as well as two more complex methods (\ac{MICE} and \ac{GAIN}) against data where increasing amounts are randomly set to missing. We then use the imputed datasets to estimate the redshift of the galaxies, using the \ac{kNN} and \acl{RF} \acs{ML} techniques. We find that the \ac{MICE} algorithm provides the lowest \acl{RMSE} and consequently the lowest prediction error, with the \ac{GAIN} algorithm the next best.
\end{abstract}
\acresetall

\freefootnote{$^*$ Authors also with Data61 CSIRO, NSW, Australia}

\section{Introduction}

Astronomical data sets often contain missing values. There are many different causes for the missing values: In a single survey, data can be missing due to incomplete observations, recording problems, data corruptions, instrument limitations or unfavourable observing conditions. When multiple surveys are combined together, data can be missing due to varying survey depths, or that some objects are not present in one of the surveys. Traditionally, samples with missing values are either ignored or the values replaced with the mean or minimum/maximum of the data set. The former means a large fraction of the data are often ignored in a data analysis task for astronomical data sets. The latter introduces large errors in the missing value, and consequently errors in the analysed results. 

With the growing popularity of deep learning, several deep learning methods for data imputation has been introduced in the literature~\citep{Pereira2020JAIR,Yoon2018gain,shang2017vigan,lee2019collagan} that take advantage of the generative nature of these methods. In this work, we investigate one of these methods,  \acl{GAIN}~\citep[\acused{GAIN}\acs{GAIN};][]{Yoon2018gain} --- in an initial study on standard datasets, we found \ac{GAIN} to be most computationally efficient among existing \ac{GAN} based algorithms, with only a slight degradation on performance. For comparison, we contrast two other popular \ac{ML} data imputation methods, \ac{kNN} and \acl{MICE}~\citep[\acused{MICE}\acs{MICE};][]{Buuren2000Netherlands}.

We examine the effect of missing data imputation on redshift estimation. For most aspects of science, knowledge of an astronomical source's redshift is an essential indicator of the distance and age of the source. Ideally, this redshift is measured directly using spectroscopy, however, for large astronomical surveys, spectroscopy data will not be available. Previously, \citet{Luken2021Australia}\footnote{Available as a pre-print at \url{https://luken.dev/Papers/2021_ASCOM_Estimate_Redshift.pdf}} performed an in-depth study of using two popular \ac{ML} methods to estimate the redshift of radio galaxies, where traditional redshift estimation techniques like template fitting have been shown to struggle \citep{norris_comparison_2018}. In this work, we investigate the accuracy of estimated redshift using imputed data at various missing rates, and compare their results against the estimation from non-imputed data. 

\section{Methods}
\label{sec:methods}

\subsection{Data set}

\begin{figure*}
    \centering
    \includegraphics[trim=0 0 0 0, width=\textwidth]{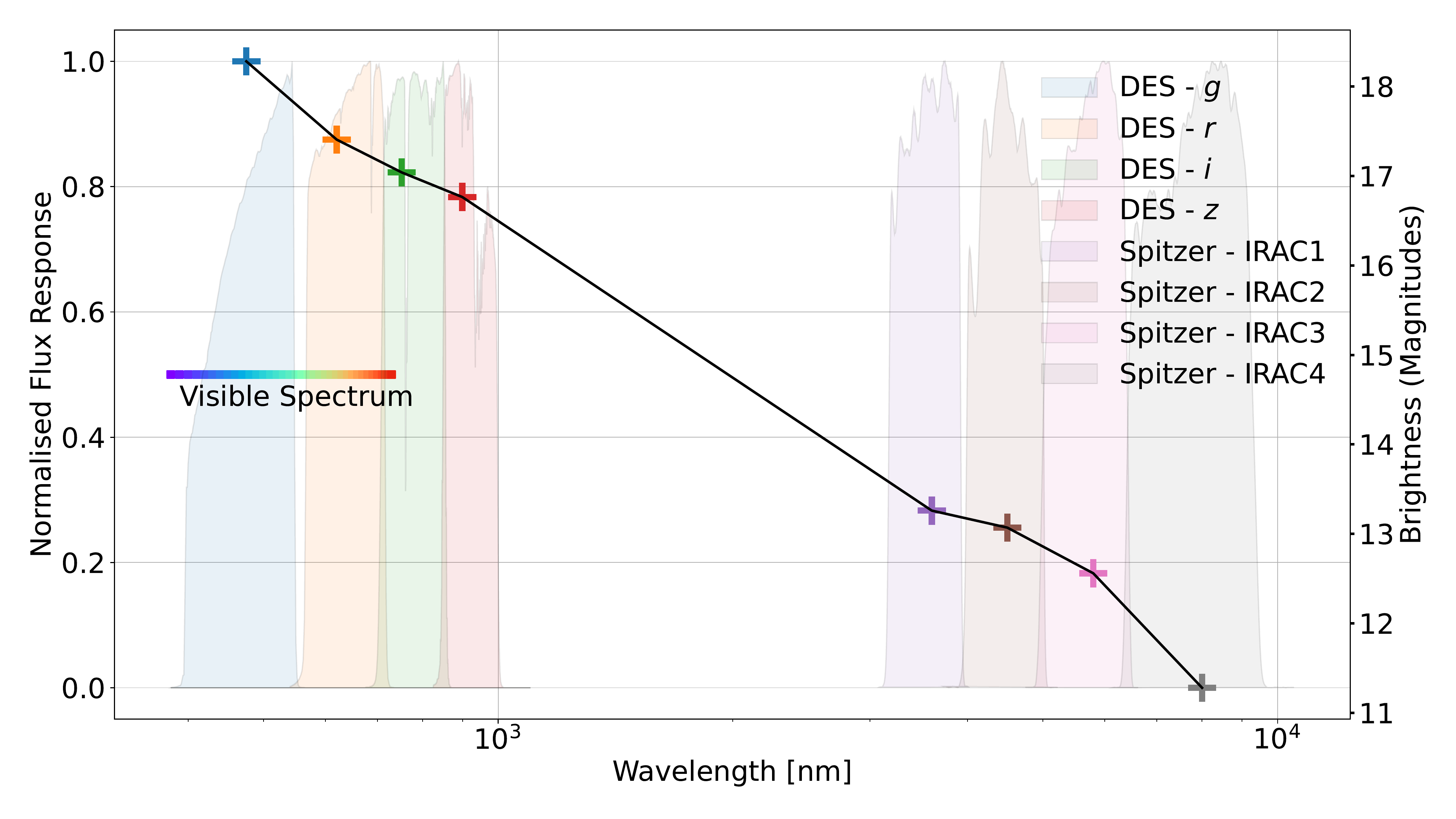}
    \caption{The \ac{SED} of the extragalactic source ATLAS3\_J033402.4-281418C, taken from the \ac{ATLAS} DR3. The background shows the filter coverage used by this work, with the ``+" in the foreground representing the measured photometry at each band. The wavelengths being measured are along the x-axis, with the left-y-axis showing the normalised flux response of the filters being measured, and the right-y-axis showing the brightness of the source in each band.
    }
    \label{fig:emu_photometry}
\end{figure*}

This work uses the same data as that described in \citep{Luken2021Australia}. The data set comprised of 1311 objects from the \acl{ATLAS} \citep[\acused{ATLAS}\acs{ATLAS};][]{norris_deep_2006,franzen_atlas_2015} radio continuum catalogue. In addition to the radio flux measured using the Australia Telescope Compact Array, the data set contains spectroscopic redshift measurements \citep{OZDES_1,OZDES_2,OZDES_3} measured primarily using the Anglo-Australian Telescope, \textit{g}, \textit{r}, \textit{i} and \textit{z} optical magnitudes measured using the Dark Energy Camera at the  Cerro Tololo Inter-American Observatory \citep{dark_energy_survey_collaboration_dark_2016}, and 3.6, 4.5, 5.4 and 8.0 $\mu$m infrared flux measurements measured using the Spitzer Space Telescope \citep{lonsdale_swire:_2003}. All attributes were standardised to $\mathcal{N}(0, 1)$. The data are available online\footnote{\url{https://github.com/kluken/Redshift-kNN-2021}, GPL-3.0 License.}, with an example \ac{SED} demonstrated in Figure~\ref{fig:emu_photometry}.

The data is partitioned randomly into two sets: using 70\% of the full data set for training, and the remaining 30\% as test data. We quantise the redshift values into 15 redshift bins for classification, with equal numbers of sources in each. The test data are used for both imputation and classification prediction. We applied different missing rates --- 2\%, 5\%, 10\% 15\%, 20\%, 25\% and 30\% --- to the test data, randomly removing a percentage of the data. The same imputed test data is then used for prediction using different classification methods. The experiment was repeated 100 times using different random seeds in order to estimate the variance in the results. This random sampling has the added effect of ensuring that the distributions of galaxies in the training and test sets are even.

\subsection{Data imputation algorithms}

The simplest way for data imputation is to replace the missing value with a value of some summary statistics. In this work, we tested imputation with \textbf{mean}, \textbf{median}, \textbf{minimum} and \textbf{maximum} values of individual attributes. We calculate the median value for a data set with even number of samples as the mean between the two middle data points.

In \textbf{\acf{kNN}} data imputation, the distance between two samples are calculated using the features without missing values. Euclidean distance was used in this work. 
The missing values are imputed using the mean value from $k$ nearest neighbours in the data set.

\textbf{\acf{MICE}} performs data imputation by filling the missing values multiple times. The algorithm initialises all missing value with the mean of their respective attribute. Each attribute's missing values are then estimated as a regression problem using the other attributes in the data set as the independent variables. The cycle is repeated multiple times. We repeated the cycle 10 times, which is the general practice. 

Finally, we tested \textbf{\acf{GAIN}}~\citep{Yoon2018gain}, where the generator's aim is imputation, and the discriminator’s goal is to distinguish between observed and imputed components. The generator is designed to maximize the discriminator’s misclassification rate, whereas the discriminator's aim is to minimize the classification loss. The \ac{GAIN} architecture also provides the discriminator with additional information in the form of \emph{hints}, which ensures that the generator generates samples according to the true underlying data distribution.

\subsection{Machine learning algorithm}
\label{sec:ml_alg}

\citet{Luken2021Australia} performed regression and classification on the data set using both \ac{kNN} with three different distance metrics and \ac{RF} algorithms. These authors concluded that for this data set, \ac{kNN} with Mahalanobis distance has the best regression and classification performances. Therefore, for this work, we concentrated our evaluation using this specific algorithm. We also evaluated the performance using \ac{RF} as a comparison. 

The \ac{kNN} algorithm \citep{cover_nearest_1967} computes a similarity matrix between all sources based on sample attributes through some distance metric. Using the similarity matrix, the \ac{kNN} algorithm finds the $k_N$ most similar sources (where $k_N$ was optimised using $k_f$-fold cross-validation), and takes either the mean value (for regression) or the mode class (for classification) of the sources' redshift as the estimated redshift for each source. 

The Mahalanobis distance metric \citep{mahalanobis1936generalized} normalises the variance and covariance of the input features by transforming the features using the inverse of the covariance matrix, $S$. The Mahalanobis distance, $d(\vec{p},\vec{q})$ between two feature vectors $\vec{p}$ and $\vec{q}$ is: 
\begin{equation} 
	\label{eqn:mahalanobis} 
    d(\vec{p},\vec{q}) = \sqrt{(\vec{p} - \vec{q})^\mathrm{T}S^{-1}(\vec{p} - \vec{q})}. 
\end{equation}

The value of $k_N$ used in the \ac{kNN} in this work is optimised using $k$-fold (where $k$ is hereafter $k_f$ and is set to 10 for this work) cross-validation. For regression, we tested all integer values of $k_N$ between 3 and 23, and for classification we tested all odd integer values between 3 and 43.

\ac{RF}~\citep{morgan_problems_1963} is an ensemble \ac{ML} method, where a set of \ac{DT} are built through bootstrpping. Each \ac{DT} is given a slightly different training data. This leads to a diversity of trees in the ensemble, which contribute to robustness of the model as a whole. The optimum number of trees is determined through $k$-fold cross-validation.

\subsection{Performance metrics}

The error metric used for imputation performance is \ac{RMSE}. To calculate the \ac{RMSE}, we mask out any data points in the test data set that is not a missing value, and only calculate the difference between the predicted value and the actual value. 

The primary error metric for the redshift estimation is the $\eta_{0.15}$ outlier rate:
\begin{equation}
    \label{eqn:outlier}
    \eta_{0.15} = \frac{\text{count}(|z_{spec} - z_{photo}| > 0.15 \times (1 + z_{spec}) )}{\text{Number Of Sources}},
\end{equation} 
where $z_{spec}$ is the estimated redshift and $z_{photo}$ is the measured redshift of a source. The $\eta_{0.15}$ outlier rate is a percentage representing the number of `catastrophic failures', and is commonly found in literature \citep[][and references therein]{Luken2021Australia}.

\subsection{Software}

This work makes use of the \emph{Scikit-learn} Python package \citep{pedregosa_scikit-learn:_2011} for the implementation of \ac{kNN} algorithms. We also used 
\ac{MICE}\footnote{\url{https://github.com/farrajota/benchmark_mice_algorithms}, MIT License.} and \ac{GAIN}\footnote{\url{https://github.com/jsyoon0823/GAIN}, Apache License, Version 2.0.} implementations from GitHub.  
The code used in this work is available on Github\footnote{\url{https://github.com/kluken/Redshift_Imputation}, GPL-3.0 License.}

\section{Results}
\label{sec:results}

\begin{figure*}
    \centering
    \includegraphics[trim=0 0 0 0, width=\textwidth]{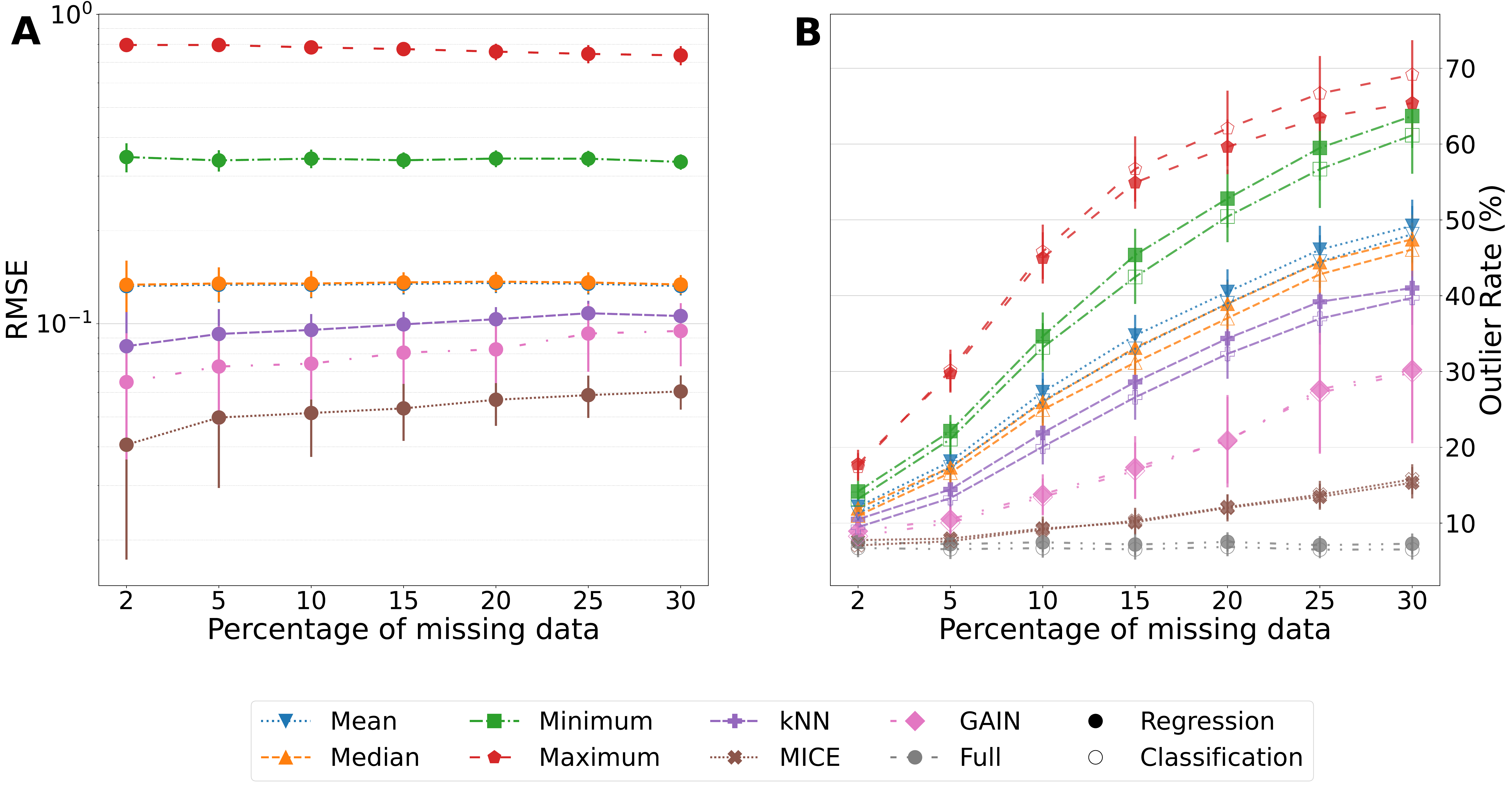}
    \caption{
        Results at different percentages of missing data. 
        (A) The \ac{RMSE} of each of the imputation methods. The y-axis is in log scale to emphasis the different results. (B) The outlier rate (defined in Equation~\ref{eqn:outlier}) of the \ac{kNN} Regression (filled markers) and Classification (hollow markers), tested on the different datasets created by the different imputation methods on the test data. Also shown on B is the outlier rate using the test set without missing data.
        All markers are set to the mean of the test, with error bars representing $\pm$1 standard deviation. 
    }
    \label{fig:outlier}
\end{figure*}

\subsection{Imputation Results}
\label{sec:results_impute}

Error rates of the imputed datasets at different missing rates are shown in Figure\ref{fig:outlier}A. 
All tests show the same trend over the different missing rates. This is due to the training method utilised -- the training/test split was completed before the test data were blanked. Therefore, the training data remains the same for all tests, ensuring that differences in the final prediction errors quoted are entirely due to the missing data in the test sample, and not potentially contaminated by errors in the training set as well. 

The \ac{MICE} algorithm performs best (\ac{RMSE} $\approx 0.05 \pm 0.01$) across all missing rates, followed by the \ac{GAIN} (\ac{RMSE} $\approx 0.08 \pm 0.02$) and \ac{kNN} (\ac{RMSE} $\approx 0.1 \pm 0.01$) algorithms. The simple Mean (\ac{RMSE} $\approx 0.13 \pm 0.01$), Median (\ac{RMSE} $\approx 0.14 \pm 0.01$), Minimum (\ac{RMSE} $\approx 0.34 \pm 0.02$) and Maximum (\ac{RMSE} $\approx 0.77 \pm 0.04$) imputation all perform progressively worse. 

The opposite was true for the time taken\footnote{All tests were completed on a desktop computer with an AMD Ryzen 7 1700 CPU and 16GB RAM}, where at 15\% missing data rate, the Mean, Median, Minimum and Maximum were imputed effectively instantly, followed by the \ac{kNN} algorithm ($\approx 1.69 \pm 1.12$ s), \ac{GAIN} algorithm ($\approx 123.86 \pm 11.55$ s), and \ac{MICE} algorithm ($\approx 871.6 \pm 202.68$ s). 

\subsection{Prediction Results}
\label{results_prediction}

Figure~\ref{fig:outlier}B (Right) shows the prediction error (the Outlier Rate, defined in Equation~\ref{eqn:outlier}) using \ac{kNN}. Results using \ac{RF} have similar trend, hence omitted here for brevity. At low missing rates (2--5\%), the outlier rates are similar to the baseline using imputed data from \ac{MICE} and \ac{GAIN}, although higher levels of missing data result in higher outlier rates.

The prediction error follows the imputation results -- the lowest outlier rates are achieved using datasets where the imputation methods achieved the lowest \ac{RMSE}. In all cases, the classification outlier rates very closely followed the regression outlier rates, with the classification outlier rates often being marginally lower, although well within the error bars. 

Looking at classification-based models predicting the datasets where 15\% of the test set was set to missing, which has a baseline of 6.54$\pm$1.36\%. The \ac{MICE}-imputed dataset performed best (10.31$\pm$1.72\%), followed by the \ac{GAIN}-imputed dataset (16.9$\pm$3.72\%) and \ac{kNN}-imputed dataset (26.6$\pm$2.95\%). As with the imputation results, the Median (31.19$\pm$2.84\%), Mean (32.99$\pm$2.85\%), Minimum (42.46$\pm$3.56\%) and Maximum (56.72$\pm$4.31\%) datasets performed the worst.

\section{Conclusion and Future Work}
\label{sec:conclusion}

Astronomical surveys are of most use when their catalogues are cross-matched with other surveys at different wavelengths. However, most astronomical surveys contain missing data for a variety of reasons, including brightness limits or issues/errors recording data for some reason. While the former is common, this work only looks at the case where data is missing at random -- where the data has not been able to be recorded correctly. Future work will look at the case where astronomical sources are too faint to be seen at particular wavelengths, and will look at including additional data where the redshift is not known. 

This work finds that in this case, the \ac{MICE} algorithm performs best at recovering the missing values, resulting in both a lower \ac{RMSE} and final outlier rate. Using the \ac{MICE} algorithm results in acceptable redshift estimates (where $\approx10\%$ of the estimates are defined as outliers) where up to 15\% of the data is missing. The \ac{GAIN} algorithm has potential with slightly higher prediction errors but at 1/7 the running time for the imputation. However, it likely under-performed due to the small sample size. The \ac{kNN} algorithm and other simple methods performed poorly in both imputation error and most importantly redshift estimation error, and should not be used going forward.

Future work will include incorporating alternative training methods, including training on imputed data, and training including astronomical sources without a measured redshift, and testing the imputation methods on datasets containing real missing values.

\acksection
The Australia Telescope Compact Array is part of the Australia Telescope National Facility which is funded by the Australian Government for operation as a National Facility managed by CSIRO. We acknowledge the Gomeroi people as the traditional owners of the Observatory site.

Based in part on data acquired at the Anglo-Australian Telescope. We acknowledge the traditional owners of the land on which the AAT stands, the Gamilaroi people, and pay our respects to elders past and present.

This work is based on archival data obtained with the Spitzer Space Telescope, which was operated by the Jet Propulsion Laboratory, California Institute of Technology under a contract with NASA. Support for this work was provided by an award issued by JPL/Caltech

This project used public archival data from the Dark Energy Survey (DES). Funding for the DES Projects has been provided by the U.S. Department of Energy, the U.S. National Science Foundation, the Ministry of Science and Education of Spain, the Science and Technology Facilities Council of the United Kingdom, the Higher Education Funding Council for England, the National Center for Supercomputing Applications at the University of Illinois at Urbana-Champaign, the Kavli Institute of Cosmological Physics at the University of Chicago, the Center for Cosmology and Astro-Particle Physics at the Ohio State University, the Mitchell Institute for Fundamental Physics and Astronomy at Texas A\&M University, Financiadora de Estudos e Projetos, Funda{\c c}{\~a}o Carlos Chagas Filho de Amparo {\`a} Pesquisa do Estado do Rio de Janeiro, Conselho Nacional de Desenvolvimento Cient{\'i}fico e Tecnol{\'o}gico and the Minist{\'e}rio da Ci{\^e}ncia, Tecnologia e Inova{\c c}{\~a}o, the Deutsche Forschungsgemeinschaft, and the Collaborating Institutions in the Dark Energy Survey.

The Collaborating Institutions are Argonne National Laboratory, the University of California at Santa Cruz, the University of Cambridge, Centro de Investigaciones Energ{\'e}ticas, Medioambientales y Tecnol{\'o}gicas-Madrid, the University of Chicago, University College London, the DES-Brazil Consortium, the University of Edinburgh, the Eidgen{\"o}ssische Technische Hochschule (ETH) Z{\"u}rich,  Fermi National Accelerator Laboratory, the University of Illinois at Urbana-Champaign, the Institut de Ci{\`e}ncies de l'Espai (IEEC/CSIC), the Institut de F{\'i}sica d'Altes Energies, Lawrence Berkeley National Laboratory, the Ludwig-Maximilians Universit{\"a}t M{\"u}nchen and the associated Excellence Cluster Universe, the University of Michigan, the National Optical Astronomy Observatory, the University of Nottingham, The Ohio State University, the OzDES Membership Consortium, the University of Pennsylvania, the University of Portsmouth, SLAC National Accelerator Laboratory, Stanford University, the University of Sussex, and Texas A\&M University.

Based in part on observations at Cerro Tololo Inter-American Observatory, National Optical Astronomy Observatory, which is operated by the Association of Universities for Research in Astronomy (AURA) under a cooperative agreement with the National Science Foundation.

\bibliographystyle{apalike}
\bibliography{references}

\begin{thebibliography}{}

\bibitem[{Childress} et~al., 2017]{OZDES_2}
{Childress}, M.~J., {Lidman}, C., {Davis}, T.~M., {Tucker}, B.~E., {Asorey},
  J., {Yuan}, F., {Abbott}, T.~M.~C., {Abdalla}, F.~B., {Allam}, S., {Annis},
  J., {Banerji}, M., {Benoit-L{\'e}vy}, A., {Bernard}, S.~R., {Bertin}, E.,
  {Brooks}, D., {Buckley-Geer}, E., {Burke}, D.~L., {Carnero Rosell}, A.,
  {Carollo}, D., {Carrasco Kind}, M., {Carretero}, J., {Castander}, F.~J.,
  {Cunha}, C.~E., {da Costa}, L.~N., {D'Andrea}, C.~B., {Doel}, P., {Eifler},
  T.~F., {Evrard}, A.~E., {Flaugher}, B., {Foley}, R.~J., {Fosalba}, P.,
  {Frieman}, J., {Garc{\'\i}a-Bellido}, J., {Glazebrook}, K., {Goldstein},
  D.~A., {Gruen}, D., {Gruendl}, R.~A., {Gschwend}, J., {Gupta}, R.~R.,
  {Gutierrez}, G., {Hinton}, S.~R., {Hoormann}, J.~K., {James}, D.~J.,
  {Kessler}, R., {Kim}, A.~G., {King}, A.~L., {Kovacs}, E., {Kuehn}, K.,
  {Kuhlmann}, S., {Kuropatkin}, N., {Lagattuta}, D.~J., {Lewis}, G.~F., {Li},
  T.~S., {Lima}, M., {Lin}, H., {Macaulay}, E., {Maia}, M.~A.~G., {Marriner},
  J., {March}, M., {Marshall}, J.~L., {Martini}, P., {McMahon}, R.~G.,
  {Menanteau}, F., {Miquel}, R., {Moller}, A., {Morganson}, E., {Mould}, J.,
  {Mudd}, D., {Muthukrishna}, D., {Nichol}, R.~C., {Nord}, B., {Ogando},
  R.~L.~C., {Ostrovski}, F., {Parkinson}, D., {Plazas}, A.~A., {Reed}, S.~L.,
  {Reil}, K., {Romer}, A.~K., {Rykoff}, E.~S., {Sako}, M., {Sanchez}, E.,
  {Scarpine}, V., {Schindler}, R., {Schubnell}, M., {Scolnic}, D.,
  {Sevilla-Noarbe}, I., {Seymour}, N., {Sharp}, R., {Smith}, M.,
  {Soares-Santos}, M., {Sobreira}, F., {Sommer}, N.~E., {Spinka}, H.,
  {Suchyta}, E., {Sullivan}, M., {Swanson}, M.~E.~C., {Tarle}, G., {Uddin},
  S.~A., {Walker}, A.~R., {Wester}, W., and {Zhang}, B.~R. (2017).
\newblock {OzDES multifibre spectroscopy for the Dark Energy Survey: 3-yr
  results and first data release}.
\newblock {\em \mnras}, 472(1):273--288.

\bibitem[Cover and Hart, 1967]{cover_nearest_1967}
Cover, T. and Hart, P. (1967).
\newblock Nearest neighbor pattern classification.
\newblock {\em IEEE Transactions on Information Theory}, 13(1):21--27.

\bibitem[{Dark Energy Survey Collaboration} et~al.,
  2016]{dark_energy_survey_collaboration_dark_2016}
{Dark Energy Survey Collaboration}, Abbott, T., Abdalla, F.~B., Aleksić, J.,
  Allam, S., Amara, A., Bacon, D., Balbinot, E., Banerji, M., Bechtol, K.,
  Benoit-Lévy, A., Bernstein, G.~M., Bertin, E., Blazek, J., Bonnett, C.,
  Bridle, S., Brooks, D., Brunner, R.~J., Buckley-Geer, E., Burke, D.~L.,
  Caminha, G.~B., Capozzi, D., Carlsen, J., Carnero-Rosell, A., Carollo, M.,
  Carrasco-Kind, M., Carretero, J., Castander, F.~J., Clerkin, L., Collett, T.,
  Conselice, C., Crocce, M., Cunha, C.~E., D'Andrea, C.~B., da~Costa, L.~N.,
  Davis, T.~M., Desai, S., Diehl, H.~T., Dietrich, J.~P., Dodelson, S., Doel,
  P., Drlica-Wagner, A., Estrada, J., Etherington, J., Evrard, A.~E., Fabbri,
  J., Finley, D.~A., Flaugher, B., Foley, R.~J., Fosalba, P., Frieman, J.,
  García-Bellido, J., Gaztanaga, E., Gerdes, D.~W., Giannantonio, T.,
  Goldstein, D.~A., Gruen, D., Gruendl, R.~A., Guarnieri, P., Gutierrez, G.,
  Hartley, W., Honscheid, K., Jain, B., James, D.~J., Jeltema, T., Jouvel, S.,
  Kessler, R., King, A., Kirk, D., Kron, R., Kuehn, K., Kuropatkin, N., Lahav,
  O., Li, T.~S., Lima, M., Lin, H., Maia, M. A.~G., Makler, M., Manera, M.,
  Maraston, C., Marshall, J.~L., Martini, P., McMahon, R.~G., Melchior, P.,
  Merson, A., Miller, C.~J., Miquel, R., Mohr, J.~J., Morice-Atkinson, X.,
  Naidoo, K., Neilsen, E., Nichol, R.~C., Nord, B., Ogando, R., Ostrovski, F.,
  Palmese, A., Papadopoulos, A., Peiris, H.~V., Peoples, J., Percival, W.~J.,
  Plazas, A.~A., Reed, S.~L., Refregier, A., Romer, A.~K., Roodman, A., Ross,
  A., Rozo, E., Rykoff, E.~S., Sadeh, I., Sako, M., Sánchez, C., Sanchez, E.,
  Santiago, B., Scarpine, V., Schubnell, M., Sevilla-Noarbe, I., Sheldon, E.,
  Smith, M., Smith, R.~C., Soares-Santos, M., Sobreira, F., Soumagnac, M.,
  Suchyta, E., Sullivan, M., Swanson, M., Tarle, G., Thaler, J., Thomas, D.,
  Thomas, R.~C., Tucker, D., Vieira, J.~D., Vikram, V., Walker, A.~R.,
  Wechsler, R.~H., Weller, J., Wester, W., Whiteway, L., Wilcox, H., Yanny, B.,
  Zhang, Y., and Zuntz, J. (2016).
\newblock The {Dark} {Energy} {Survey}: more than dark energy - an overview.
\newblock {\em \mnras}, 460:1270--1299.

\bibitem[Franzen et~al., 2015]{franzen_atlas_2015}
Franzen, T. M.~O., Banfield, J.~K., Hales, C.~A., Hopkins, A., Norris, R.~P.,
  Seymour, N., Chow, K.~E., Herzog, A., Huynh, M.~T., Lenc, E., Mao, M.~Y., and
  Middelberg, E. (2015).
\newblock {ATLAS} - {I}. {Third} release of 1.4 {GHz} mosaics and component
  catalogues.
\newblock {\em \mnras}, 453:4020--4036.

\bibitem[Lee et~al., 2019]{lee2019collagan}
Lee, D., Kim, J., Moon, W.-J., and Ye, J.~C. (2019).
\newblock Collagan: Collaborative gan for missing image data imputation.
\newblock In {\em Proceedings of the IEEE/CVF Conference on Computer Vision and
  Pattern Recognition}, pages 2487--2496.

\bibitem[{Lidman} et~al., 2020]{OZDES_3}
{Lidman}, C., {Tucker}, B.~E., {Davis}, T.~M., {Uddin}, S.~A., {Asorey}, J.,
  {Bolejko}, K., {Brout}, D., {Calcino}, J., {Carollo}, D., {Carr}, A.,
  {Childress}, M., {Hoormann}, J.~K., {Foley}, R.~J., {Galbany}, L.,
  {Glazebrook}, K., {Hinton}, S.~R., {Kessler}, R., {Kim}, A.~G., {King}, A.,
  {Kremin}, A., {Kuehn}, K., {Lagattuta}, D., {Lewis}, G.~F., {Macaulay}, E.,
  {Malik}, U., {March}, M., {Martini}, P., {M{\"o}ller}, A., {Mudd}, D.,
  {Nichol}, R.~C., {Panther}, F., {Parkinson}, D., {Pursiainen}, M., {Sako},
  M., {Swann}, E., {Scalzo}, R., {Scolnic}, D., {Sharp}, R., {Smith}, M.,
  {Sommer}, N.~E., {Sullivan}, M., {Webb}, S., {Wiseman}, P., {Yu}, Z., {Yuan},
  F., {Zhang}, B., {Abbott}, T.~M.~C., {Aguena}, M., {Allam}, S., {Annis}, J.,
  {Avila}, S., {Bertin}, E., {Bhargava}, S., {Brooks}, D., {Carnero Rosell},
  A., {Carrasco Kind}, M., {Carretero}, J., {Castander}, F.~J., {Costanzi}, M.,
  {da Costa}, L.~N., {De Vicente}, J., {Doel}, P., {Eifler}, T.~F., {Everett},
  S., {Fosalba}, P., {Frieman}, J., {Garc{\'\i}a-Bellido}, J., {Gaztanaga}, E.,
  {Gruen}, D., {Gruendl}, R.~A., {Gschwend}, J., {Gutierrez}, G., {Hartley},
  W.~G., {Hollowood}, D.~L., {Honscheid}, K., {James}, D.~J., {Kuropatkin}, N.,
  {Li}, T.~S., {Lima}, M., {Lin}, H., {Maia}, M.~A.~G., {Marshall}, J.~L.,
  {Melchior}, P., {Menanteau}, F., {Miquel}, R., {Palmese}, A.,
  {Paz-Chinch{\'o}n}, F., {Plazas}, A.~A., {Roodman}, A., {Rykoff}, E.~S.,
  {Sanchez}, E., {Santiago}, B., {Scarpine}, V., {Schubnell}, M., {Serrano},
  S., {Sevilla-Noarbe}, I., {Suchyta}, E., {Swanson}, M.~E.~C., {Tarle}, G.,
  {Tucker}, D.~L., {Varga}, T.~N., {Walker}, A.~R., {Wester}, W., {Wilkinson},
  R.~D., and {DES Collaboration} (2020).
\newblock {OzDES multi-object fibre spectroscopy for the Dark Energy Survey:
  results and second data release}.
\newblock {\em \mnras}, 496(1):19--35.

\bibitem[Lonsdale et~al., 2003]{lonsdale_swire:_2003}
Lonsdale, C.~J., Smith, H.~E., Rowan-Robinson, M., Surace, J., Shupe, D., Xu,
  C., Oliver, S., Padgett, D., Fang, F., Conrow, T., Franceschini, A., Gautier,
  N., Griffin, M., Hacking, P., Masci, F., Morrison, G., O'Linger, J., Owen,
  F., Pérez-Fournon, I., Pierre, M., Puetter, R., Stacey, G., Castro, S.,
  Polletta, M. d.~C., Farrah, D., Jarrett, T., Frayer, D., Siana, B., Babbedge,
  T., Dye, S., Fox, M., Gonzalez-Solares, E., Salaman, M., Berta, S., Condon,
  J.~J., Dole, H., and Serjeant, S. (2003).
\newblock {SWIRE}: {The} {SIRTF} {Wide}-{Area} {Infrared} {Extragalactic}
  {Survey}.
\newblock {\em Publications of the Astronomical Society of the Pacific},
  115:897--927.

\bibitem[Luken et~al., 2021]{Luken2021Australia}
Luken, K.~J., Norris, R.~P., Wang, X.~R., Park, L. A.~F., and Filipovic, M.~D.
  (2021).
\newblock Estimating galaxy redshift in radio-selected datasets using machine
  learning.
\newblock "\url{https://luken.dev/Papers/2021_ASCOM_Estimate_Redshift.pdf}".
\newblock under review.

\bibitem[Mahalanobis, 1936]{mahalanobis1936generalized}
Mahalanobis, P.~C. (1936).
\newblock On the generalized distance in statistics.
\newblock In {\em On the Generalised Distance in Statistics.} National
  Institute of Science of India.

\bibitem[Morgan and Sonquist, 1963]{morgan_problems_1963}
Morgan, J.~N. and Sonquist, J.~A. (1963).
\newblock Problems in the {Analysis} of {Survey} {Data}, and a {Proposal}.
\newblock {\em Journal of the American Statistical Association},
  58(302):415--434.

\bibitem[Norris et~al., 2006]{norris_deep_2006}
Norris, R.~P., Afonso, J., Appleton, P.~N., Boyle, B.~J., Ciliegi, P., Croom,
  S.~M., Huynh, M.~T., Jackson, C.~A., Koekemoer, A.~M., Lonsdale, C.~J.,
  Middelberg, E., Mobasher, B., Oliver, S.~J., Polletta, M., Siana, B.~D.,
  Smail, I., and Voronkov, M.~A. (2006).
\newblock Deep {ATLAS} {Radio} {Observations} of the {Chandra} {Deep}
  {Field}-{South}/{Spitzer} {Wide}-{Area} {Infrared} {Extragalactic} {Field}.
\newblock {\em Astronomical Journal}, 132:2409--2423.

\bibitem[{Norris} et~al., 2019]{norris_comparison_2018}
{Norris}, R.~P., {Salvato}, M., {Longo}, G., {Brescia}, M., {Budavari}, T.,
  {Carliles}, S., {Cavuoti}, S., {Farrah}, D., {Geach}, J., {Luken}, K.,
  {Musaeva}, A., {Polsterer}, K., {Riccio}, G., {Seymour}, N.,
  {Smol{\v{c}}i{\'c}}, V., {Vaccari}, M., and {Zinn}, P. (2019).
\newblock {A Comparison of Photometric Redshift Techniques for Large Radio
  Surveys}.
\newblock {\em \pasp}, 131(1004):108004.

\bibitem[Pedregosa et~al., 2011]{pedregosa_scikit-learn:_2011}
Pedregosa, F., Varoquaux, G., Gramfort, A., Michel, V., Thirion, B., Grisel,
  O., Blondel, M., Prettenhofer, P., Weiss, R., Dubourg, V., Vanderplas, J.,
  Passos, A., Cournapeau, D., Brucher, M., Perrot, M., and Duchesnay, E.
  (2011).
\newblock Scikit-learn: {Machine} {Learning} in {Python}.
\newblock {\em Journal of Machine Learning Research}, 12:2825--2830.

\bibitem[Pereira et~al., 2020]{Pereira2020JAIR}
Pereira, R.~C., Santos, M.~S., Rodrigues, P.~P., and Abreu, P.~H. (2020).
\newblock Reviewing autoencoders for missing data imputation: Technical trends,
  applications and outcomes.
\newblock {\em Journal of Artificial Intelligence Research}, 69:1255--1285.

\bibitem[S.~van Buuren, 2000]{Buuren2000Netherlands}
S.~van Buuren, C.~O. (2000).
\newblock {\em Multivariate Imputation by Chained Equations}.
\newblock Public Health, Netherlands.

\bibitem[Shang et~al., 2017]{shang2017vigan}
Shang, C., Palmer, A., Sun, J., Chen, K.-S., Lu, J., and Bi, J. (2017).
\newblock Vigan: Missing view imputation with generative adversarial networks.
\newblock In {\em 2017 IEEE International Conference on Big Data (Big Data)},
  pages 766--775. IEEE.

\bibitem[Yoon et~al., 2018]{Yoon2018gain}
Yoon, J., Jordon, J., and Schaar, M. (2018).
\newblock {GAIN}: Missing data imputation using generative adversarial nets.
\newblock In {\em International Conference on Machine Learning}, pages
  5689--5698. PMLR.

\bibitem[{Yuan} et~al., 2015]{OZDES_1}
{Yuan}, F., {Lidman}, C., {Davis}, T.~M., {Childress}, M., {Abdalla}, F.~B.,
  {Banerji}, M., {Buckley-Geer}, E., {Carnero Rosell}, A., {Carollo}, D.,
  {Castander}, F.~J., {D'Andrea}, C.~B., {Diehl}, H.~T., {Cunha}, C.~E.,
  {Foley}, R.~J., {Frieman}, J., {Glazebrook}, K., {Gschwend}, J., {Hinton},
  S., {Jouvel}, S., {Kessler}, R., {Kim}, A.~G., {King}, A.~L., {Kuehn}, K.,
  {Kuhlmann}, S., {Lewis}, G.~F., {Lin}, H., {Martini}, P., {McMahon}, R.~G.,
  {Mould}, J., {Nichol}, R.~C., {Norris}, R.~P., {O'Neill}, C.~R., {Ostrovski},
  F., {Papadopoulos}, A., {Parkinson}, D., {Reed}, S., {Romer}, A.~K.,
  {Rooney}, P.~J., {Rozo}, E., {Rykoff}, E.~S., {Sako}, M., {Scalzo}, R.,
  {Schmidt}, B.~P., {Scolnic}, D., {Seymour}, N., {Sharp}, R., {Sobreira}, F.,
  {Sullivan}, M., {Thomas}, R.~C., {Tucker}, D., {Uddin}, S.~A., {Wechsler},
  R.~H., {Wester}, W., {Wilcox}, H., {Zhang}, B., {Abbott}, T., {Allam}, S.,
  {Bauer}, A.~H., {Benoit-L{\'e}vy}, A., {Bertin}, E., {Brooks}, D., {Burke},
  D.~L., {Carrasco Kind}, M., {Covarrubias}, R., {Crocce}, M., {da Costa},
  L.~N., {DePoy}, D.~L., {Desai}, S., {Doel}, P., {Eifler}, T.~F., {Evrard},
  A.~E., {Fausti Neto}, A., {Flaugher}, B., {Fosalba}, P., {Gaztanaga}, E.,
  {Gerdes}, D., {Gruen}, D., {Gruendl}, R.~A., {Honscheid}, K., {James}, D.,
  {Kuropatkin}, N., {Lahav}, O., {Li}, T.~S., {Maia}, M.~A.~G., {Makler}, M.,
  {Marshall}, J., {Miller}, C.~J., {Miquel}, R., {Ogando}, R., {Plazas}, A.~A.,
  {Roodman}, A., {Sanchez}, E., {Scarpine}, V., {Schubnell}, M.,
  {Sevilla-Noarbe}, I., {Smith}, R.~C., {Soares-Santos}, M., {Suchyta}, E.,
  {Swanson}, M.~E.~C., {Tarle}, G., {Thaler}, J., and {Walker}, A.~R. (2015).
\newblock {OzDES multifibre spectroscopy for the Dark Energy Survey: first-year
  operation and results}.
\newblock {\em \mnras}, 452(3):3047--3063.

\end{thebibliography}

\end{document}